 \documentstyle{article}
 \def\nat{{Nature}\ }

 \catcode`@=11
 \def\mincir{\raise -2.truept\hbox{\rlap{\hbox{$\sim$}}\raise5.truept
 \hbox{$<$}\ }}
 \def\magcir{\raise -2.truept\hbox{\rlap{\hbox{$\sim$}}\raise5.truept
 \hbox{$>$}\ }}
 \def\minmag{\raise-2.truept\hbox{\rlap{\hbox{$<$}}\raise 6.truept\hbox
 {$>$}\ }}
 \def\cross{\displaystyle / \kern-0.2truecm\hbox{$\backslash$}}
 \def\quad@rato#1#2{{\vcenter{\vbox{
	 \hrule height#2pt
	 \hbox{\vrule width#2pt height#1pt \kern#1pt \vrule width#2pt}
	 \hrule height#2pt} }}}
 \def\dall{\mathchoice
 \quad@rato5{.5}\quad@rato5{.5}\quad@rato{3.5}{.35}\quad@rato{2.5}{.25} }
 \font\s@=cmss10\font\s@b=cmbx8
 \def\real{{\hbox{\s@ l\kern-.5mm R}}}
 \def\M{{\hbox{\s@ l\kern-.5mm M}}}
 \def\K{{\hbox{\s@ l\kern-.5mm K}}}
 \def\nat{{\hbox{\s@ l\kern-.5mm N}}}
 \def\integ{{\mathchoice
  {\hbox{\s@ Z\kern-1.5mm Z}}
  {\hbox{\s@ Z\kern-1.5mm Z}}
  {\hbox{{\s@b Z\kern-1.2mm Z}}}
  {\hbox{{\s@b Z\kern-1.2mm Z}}}  }}
 \def\compl{{\hbox{\s@ C\kern-1.7mm\raise.4mm\hbox{\s@b l}\kern.8mm}}}
 \def\toro{{\hbox{\s@ T\kern-1.9mm T}}}
 \def\unity{{\hbox{\s@ 1\kern-.8mm l}}}
 \font\bold@mit=cmmib10
 \def\setbmit{\textfont1=\bold@mit}
 \def\bmit#1{\hbox{\textfont1=\bold@mit$#1$}}
 \catcode`@=12
 \def\gsim{\ \rlap{\raise 2pt \hbox{$>$}}{\lower 2pt \hbox{$\sim$}}\ }
 \def\lsim{\ \rlap{\raise 2pt \hbox{$<$}}{\lower 2pt \hbox{$\sim$}}\ }
 \def\BK*gam{B\rightarrow K^*\gamma}

 \def\k0{K^+ \rightarrow \pi^0 e^+ {\nu}_e}

 \def\ax{{\tilde a} }

 \newskip\humongous \humongous=0pt plus 1000pt minus 1000pt

 \newif\ifdtup

\begin{document}

 \centerline{\bf Mixed dark matter models with a non--thermal }

 \centerline{\bf hot component: fluctuation evolution.}

 \medskip\medskip\medskip

 \rm

 E. Pierpaoli

 S.A.~Bonometto

 \medskip

 \it
 Dipartimento di Fisica dell'Universit\'a di Milano,
 Via Celoria 16, I-20133 Milano, Italy

 I.N.F.N. -- Sezione di Milano

 \rm

 \vglue 2.8truecm
 \noindent
 {\bf Abstract} -- We calculate the linear evolution for a class of mixed dark
 matter models, where the hot component
 derives from the  decay of a heavier particle (CntHDM models). These
 models differ from ordinary mixed models based on massive neutrinos (CHDM
 models), for which the hot component has a phase space distribution which
 derives from a thermal one. In CntHDM models the density of the hot component
 and the derelativisation redshift of its quanta are indipendent parameters.
 In this work we provide
 the spectra for
 a number of CntHDM models,
 and compare them with CDM and CHDM spectra.
 If PQ symmetry and SUSY simultaneously hold,
 the lightest standard neutralino can be expected to decay into axino
 and photon.
 We briefly summarise the features of this
 particle model which gives rise
 to the cosmological framework discussed here,
 as a fairly generic consequence.
Other frameworks which lead to similar models are also briefly discussed.
\vskip 1.5truecm
\noindent
key words: Cosmology -- galaxies : origin.

 \vfill\eject
 \noindent
 {\bf 1. Introduction. }

 \smallskip
 \noindent
 Mixed dark matter (MDM) models (Valdarnini $\&$
 Bonometto 1985, Bonometto $\&$ Valdarnini 1985, Achilli et al. 1985, Holtzman
 1989, see also Fang, Li, $\&$ Xiang 1984) allow a better fit of large scale
 data than most other models (Schaefer, Shafi, \& Stecker 1989; van Dalen \&
 Schaefer 1992; Schaefer $\&$ Shafi 1992; Taylor $\&$ Rowan--Robinson 1992;
 Davis, Summers, $\&$ Schlegel 1992, Holtzman $\&$ Primack 1993; Pogosyan $\&$
 Starobinsky 1993; Liddle $\&$ Lyth 1993, Klypin et al. 1993, Klypin,
Nolthenius
 $\&$ Primack 1993, Bonometto et al. 1994). Recent treatments of MDM were
mostly
 based on a mixture of CDM, HDM and baryons with $\Omega_c/\Omega_h/\Omega_b =
 0.6/0.3/0.1$ ($\Omega$: density parameters), although different ratios were
 also considered. If HDM is made of fermions of mass $m_h$, with $g_f$
 spin states,  which decoupled when $T \gg m_h c^2$,
 it is $\Omega_h \simeq 0.3(m_h g_f/14\, {\rm
 eV})$. The standard candidate as HDM component is the $\tau$--neutrino,
assumed
 to have a mass $m_h = 7\, $eV.

Hereafter this version of MDM will be called CHDM.  Within its context
$\Omega_h$ and the derelativization redshift  for the hot component $z_{de}
\simeq  m_h c^2 /3\, T_{o,\nu}$  are  connected through $m_h$ ($T_{o,\nu}$ is
the present neutrino  temperature).  In turn, $z_{de}$ fixes the
mass scale $M_{D,h}$ above which fluctuations  in the hot component are not
erased by free streaming after their entry  in the horizon.

Instead of being composed of massive neutrinos, the hot component could have
arisen from the decay  of heavier particles. In this case, the link among
$m_h$, $g_f$, $\Omega_h$,  $z_{de}$, and $M_{D,h}$ is not so cogent. There are
still precise constraints due, e.g., to primeval nucleosynthesis,
but  the parameter
space is spanned by two independent continuous  parameters (e.g., $\Omega_h$
and $z_{de}$), instead of  a continuous and a discrete parameter (e.g.,
$\Omega_h$ and $g_f$). The distribution  of hot particles in
momentum space is then quite different from the one  originating from a thermal
distribution.  In this note we shall discuss  such alternative to CHDM
(hereafter denominated CntHDM).

 The decay of mother particles takes place when fluctuations over the relevant
 scales are still well outside the horizon. Daugther particles are then
 relativistic and an analytical treatment of the evolution of fluctuations
 during these stages can be given. The critical events to set up the shape of
 the final spectrum are the entry of the fluctuations in the horizon and the
 derelativization of the particles forming its hot component. These two events
 can take place in different order, according to the scale considered and to
the
 model parameters. At a redshift $\sim 10^3$, primeval plasma (re)combines. In
 purely baryonic models recombination is a critical step, as it sets up a
 minimal mass scale for the survival of baryonic fluctuations and provides
 information on very--small--scale CBR fluctuations. In models dominated by
 DM, these stages are not so essential, as baryons will later fall in potential
 wells created by CDM and/or HDM, while the relevant scales for CBR fluctuation
 observations however enter the horizon after recombination. The presence
 of radiation and baryons is critical for fluctuations entering the horizon
 before equivalence. The growing of the CDM component is then almost stopped by
 Meszaros effect, i.e. by the fact that baryon and radiation can only oscillate
 as sound waves while, until equivalence, they are the dominant {\sl substance}
 in the Universe. After equivalence, as is known, DM fluctuation growing can
 restart.

 Furthermore, the baryonic component has a  non neglegible influence on the
detailed  final spectrum. Our models take also into account a massless neutrino
 background and its fluctuations (which are rapidly dissipated at the entry in
the horizon). Three, two or one massless neutrinos were considered, according
to the  model parameters, as is discussed below.

 The main output of this paper are a number of spectra for CntHDM models,
 computed in detail through all their linear evolution, down to
 $z = 0$. The models considered are given in Table 1 of sec.~2. As
 is also outlined in sec.~4, the study of each model still requires a
 fairly large amount of CPU, and further numerical analysis is
 still in progress.

 A particle model, recently discussed by Bonometto, Gabbiani $\&$
 Masiero (1994, BGM), originates CntHDM
 models as a fairly generic consequence.
 We shall briefly outline some features of this particle model
 in sec.~2. The class of cosmological models
 described here is however relevant, also
 independently of that specific particle physics.  In sec.~3 we will study the
 analytical part of the evolution of density fluctuations. In sec.~4 the
 numerical treatment will be described. Sec.~5 is devoted to a discussion
 of the results.

 \medskip
 \noindent
 {\bf 2. A particle model. }

 \vglue 0.1truecm
 \noindent
 In this section we debate a specific particle model which gives rise to a
 cosmological framework of the kind discussed in the next sections.
 Although this model is complitely consistent, the cosmological results
 have a validity which extends well beyond this peculiar particle physics.
 In recent years cosmologists have become familiar with supersymmetry (SUSY)
 and Peccei--Quinn (PQ) symmetry. Both of them provide CDM candidates. The
 lightest mass eigenstate of neutral fermion SUSY partners, called {\sl
 neutralino} ($\chi$), has a mass $m_\chi \magcir 30\, $GeV and must therefore
 decouple when $T$ is significantly below $m_\chi$ (Lee $\&$ Weinberg 1977,
 Ellis et al., 1984). The goldstone boson associated to PQ symmetry, arising
 at $T < V_{PQ}$ (scale of PQ symmetry breakdown), is called axion ($a$). If
 both SUSY and PQ symmetry are implemented, also SUSY partners of $a$'s exist;
 the fermion partner of $a$ is called axino ($\tilde a$). In BGM it is shown
 that we can expect two $\tilde a$ background to exist. Besides thermal $\ax
 $'s, decoupling slightly below $V_{PQ}$, probably diluted by entropy inflow(s)
 at later phase transition(s), non--thermal $\ax$'s originate from $\chi$
decay.
 The former component is cold, the latter one is a non--thermal hot component.
 It is possible that all DM consists of $\ax$'s, which can account for both
 components of MDM.

 The decay $\chi \to \ax + \gamma$ takes place at $z_{\chi,Dy} \sim 10^9$ and
 $\ax$'s have initially momentum $P = m_\chi c/2$. They derelativize at $z_{de}
 \simeq z_{\chi,Dy} 2 m_\ax / m_\chi $. The outcoming density parameter
 $\Omega_\ax = \Omega_h$ is linked to the axino mass ($m_\ax$) and $\chi$
number
 density at the eve of their decay ($n_\chi$). These quantities are linked by a
 number of particle relations, discussed in BGM and summarized in fig.~1. In
 such figure $m_{sf}$ is the scale of the soft SUSY breaking ({\sl sfermion}
 scale.

 Different curves refer to possible choices of $m_\chi$ and $m_t$
 (top quark mass), as is detailed in its caption. For each value of
 $m_{sf}$, fig.~1 provides the values allowed for the axino mass $m_{\tilde a}$
 (decreasing curves) and for the quantity
 $z_{de} V_{PQ} / \Omega_{\tilde a}$, where the particle parameter
 $V_{PQ}$, expected to range between 5$\times 10^9$GeV and 10$^{12}$GeV, is
 combined with $z_{de}$ ($\tilde a$ derelativization redshift)
 and $\Omega_{\tilde a}$ (hot DM density parameter).

 Fig.~1 shows that the quantity
 $z_{de} V_{PQ} / \Omega_{\tilde a}$
 can be expected to range between $10^{13}$GeV and $10^{16}$GeV,
 for reasonable values of $m_{sf}$. In turn, it is possible to find
 various combinations of particle parameters which give $\Omega_{\tilde a}$
 in the range 0.1--1, and $z_{de}$ in the range 10$^3$--10$^5$.

 Further constraints are set by the number of massless neutrinos still present
 during primeval nucleosynthesis. Let $ g_\nu$ be the number of massless $\nu$
 spin states present during the  nucleosynthesis ($\nu$'s with mass much below
 nucleosynthesis temperatures are to be considered massless).
 Standard stellar abundance data for light elements allow
 up to $g_\nu \simeq 7$. Accordingly, if $g_\nu = 6$ (3 standard neutrinos)
 we are allowed $g^+ = 1$ extra spin states. It can be  $g^+ = 3$ or
 $g^+ = 5$, if one or two neutrino(s) are supposed to be quite heavy and
 to have already decayed before nucleosynthesis.
 Neutrinos with a mass $m_\nu =$7 eV ($=8
 \times 10^4$K) derelativize at a redshift $z \simeq
 1.5 \times 10^4$ and yield $\Omega_\nu = 0.3$. A particle
 allowed half of their energy density at nucleosynthesis and derelativizing at
 $z_{de}$ will then yield $\Omega_h \simeq z_{de} / 10^5$. If the number of
 extra neutrino spin states allowed during nucleosynthesis is $g^+$, the
 limit for $\Omega_h$ reads:
 $$
 \Omega_h \mincir g^+\, z_{de} / 10^5 ~.
 \eqno (2.1)
 $$
 The number $g_\nu$ of effective massless neutrinos is used in the
 equations ruling the expansion of the Universe through the
 quantity $w_\nu = (7/4)(4/11)^{4 \over 3} g_\nu$, that will be
 used below.

 For all our models we took $\Omega_{total} = 1$, $\Omega_b = 0.05$,
 and H = 50$\, $km$\, $s$^{-1}$Mpc$^{-1}$. Values of $\Omega_h$ and
 $z_{de}$ ranging in the intervals 0.15--0.30 and $4 \times 10^3$--$
 6 \times 10^4$ were considered. Models with low $z_{de}$ show interesting
 features and the parameter space was explored in that region
 in more detail.
 In Table 1 we report the values of $\Omega_h$, $z_{de}$, and
 the number of massless neutrinos, for those models
 whose spectra are worked out here.

%

 \smallskip
 \begin{table}[tp]
 \centering
 \caption[]{Parameters for  CntHDM models}
 \tabcolsep 7pt
 \begin{tabular}{lccccccccccc} \\ \\
 & {\rm model} & 1 & 2 & 3 & 4 & 5 & 6 & 7 & 8 & 9 & 10 \\
 & $\Omega_{\tilde a}$ & 0.30 & 0.30 & 0.30 & 0.15  & 0.20  & 0.15 & 0.20
 & 0.20 & 0.15 & 0.20  \\
 & $z_{de}/10^4$ & 1.5 & 6 & 3 & 1.5  & 0.7  & 0.7  & 0.9 & 1 & 0.4 & 0.4 \\
 & {\rm massless~}$\nu$'{\rm s} & 2 & 3 & 3 & 3  & 2  & 2  & 2 & 1 & 2 & 1 \\
 \\
 \end{tabular}
 \end{table}

  \medskip
 \noindent
 {\bf 3. Density fluctuation evolution: analytical aspects. }

 \vglue 0.1truecm
 \noindent
 Let $f(x_\alpha,P_\alpha,t) $ be the ntH (non--thermal hot) $\ax$ distribution
 in phase space ($x_\alpha$, $P_\alpha = P\, n_\alpha$:
 particle coordinates and
 momentum). We must also consider the source of $\ax$'s due to $\chi$--decay:
 $S(x_\alpha,P_\alpha,t) $. Up to first order,
 $$
 f = f_o(P,t) [1 + \tilde \epsilon(x_\alpha,P,t) ]~,~~~
 S = S_o(P,t) [1 + \sigma_s(x_\alpha,P,t) ]
 \eqno (3.1)
 $$
(Notice that $\tilde \epsilon$ and $\sigma_s$ do not depend on $n_{\alpha}$).
 The analysis is based on the equation
 $$
 S = \partial f / \partial t +
 (\partial f / \partial x_\alpha) \dot x_\alpha +
 (\partial f / \partial P) \dot P
 \eqno (3.2)
 $$
 taking into account that $\dot x_\alpha = cn_\alpha (P/P_o) (T_o/T),~
 \dot P = P[(\dot T/T) + y/4]$. Here $P_o c$ is energy, $T$ is radiation
 temperature, $y$ yields perturbation self--gravity and will be given in
 eq.(3.7) below. At the 0--th order,
 eq.~(3.2) yields
 $$
 S_o = \partial f_o / \partial t + P (\dot T/ T)(\partial f_o
 / \partial P) ~~,
 \eqno (3.3)
 $$
 while the source term, due to $\chi$--decay, is
 $$
 S_o = (h^3/4\pi) n_{\chi,dg} (T/T_{\chi,Dy} )^3 \exp(-t/t_{\chi,Dy} ) \,
 (t_{\chi,Dy} P^2)^{-1} \delta(m_{\chi,Dy} c/2 - P)
 \eqno (3.4)
 $$
($t_{\chi,Dy} $: $\chi$ mean life, $n_{\chi,dg}$:$\chi$\'s number density
at the decoupling ).
With such source terms the
 solution of eq.~(3.3) reads
 $$
 f_o = (h^3/4\pi)  n_{\chi,dg} \exp(-Q) (2Q/P^3) \, \theta(m_{\chi,Dy} c/2 - P)
 \eqno (3.5)
 $$
 with
 $$
 Q = (2P/m_\chi c)^2 (t/t_{\chi,Dy} )
 \eqno (3.6)
 $$
 (notice that $Q\, \delta(m_\chi c/2 - P) = t/t_{\chi,Dy} $);
 $\chi$ decays occur in radiation dominated era, when $t/t_{\chi,Dy}
 = (T_{\chi,Dy} / T)^2 $.

 We shall treat the first order equation for the case of a plane
 wave in $x_3$ direction. In such direction the component of $P$
 is $\mu\, P$. Let also $h_{\alpha \beta} = \eta_{\alpha \beta}
 - g_{\alpha \beta} $ express the deviation of the metric tensor
 $g_{\alpha \beta}$ from purely spatially flat $\eta_{\alpha \beta}$
 components (${\alpha,\, \beta} = 1..3$; $t$ being the universal time,
 only $g_{oo}$ does not vanish, besides of $g_{\alpha \beta}$).
 Then
 $$
 y = (1-\mu^2) 3 \dot h_t - (1-3\mu^2) \dot h_{33}
 = 2P_o(\mu) \dot h_t + 2P_2 (\mu) \dot h_3 ~,
 \eqno (3.7)
 $$
 where
 $$
 3h_t = \sum_{\alpha = 1}^3 h_{\alpha \alpha} ~, ~~~~~~~
 h_3 =
 h_t - h_{33}~~.
 \eqno (3.8)
 $$
 Let us then take
 $ {\tilde \epsilon} = \epsilon (k,P,\mu,t) \exp(ikx_3T/T_o)$, in order
 to treat each length scale
 separately ($\rho_c$: present critical density; $P_l$: Legendre
 Polynomials).

 The first order equation splits in two parts. Let be $K = kcT/T_o$.
 For $P < m_\chi c/2$,
 $$
 \dot \epsilon + i \epsilon \mu K P/P_o = (1 + 2Q) y/4 ~,
 \eqno (3.9)
 $$
 while, for $P = m_\chi c/2$, it is $\epsilon = \sigma_s + yt/2$.
 The latter equation holds at $\chi$ decay and sets initial conditions to
 eq.~(3.9). Such equation is to be solved for any $\mu$ and $P$; however
 $P$ is related to
 $Q$ (see 3.6) and the
distribution  $f_o$ decreases exponentially with $Q$
 (see 3.5). This will allow to select a finite set of $Q$ values.
 As far as $\mu$ is concerned, we perform the expansion
 $$
 \epsilon (k,P,\mu,t) = \sum_{l=0}^\infty (-i)^l \sigma_l (k,P,t)
 P_l (\mu) ~.
 \eqno (3.10)
 $$
 Then, taking eq.~(3.7) into account, eq.~(3.9) yields the system
 $$
 \dot \sigma_o = -(1/3)K(P/P_o) \sigma_1 + (1/2+Q) \dot h_t
 ~,~~
 \dot \sigma_1 = K(P/P_o) (\sigma_o - 2\sigma_2/5)
 $$
 $$
 \dot \sigma_2 = K(P/P_o) (2\sigma_1/3 - 3\sigma_3/7) + (1/2+Q)
 \dot h_3 ~,~~
 \dot \sigma_l = K(P/P_o) (l_-\sigma_{l-1} - l_+\sigma_{l+1})
 ~~~~ (l>2)
 \eqno (3.11)
 $$
[here $l_- = l/(2l-1),~l_+ = (l+1)/(2l+3)$].

 Radiation is to be treated very much alike. The critical difference is that
 photons are massless and therefore all $P$ behave in the same way
(furthermore,
 of course, the $P$ distribution is different). Integrating the photon
 distribution over $P$ and expanding the outcoming $\delta (k,\mu,t) =
 \sum_{l=0}^\infty (-i)^l \delta_l (k,t) P_l (\mu) $, we obtain a closed system
 analogous to (3.11), which reads
 $$
 \dot \delta_o = -(1/3)K \delta_1 + 2\dot h_t
 ~,~~
 \dot \delta_1 = K (\delta_o - 2\delta_2/5) - (\delta_1-4w) \nu
 $$
 $$
 \dot \delta_2 = K (2\delta_1/3 - 3\delta_3/7) +
 2\dot h_3 - \delta_2 \, 9\nu/10 ~,~~
 \dot \delta_l = K (l_-\delta_{l-1} - l_+\delta_{l+1}) - \delta_l \nu
 ~~~~ (l>2).
 \eqno (3.12)
 $$
 Apart of the replacement $1/2+Q \to 2$, the main difference are the
 terms accounting for matter--radiation collisions: $\nu = n_e \sigma_T c$
 is the inverse $\gamma$ collision time ($n_e$: electron number density,
 $\sigma_T$: Thomson cross--section) and $w$ is the velocity field in baryons.
 Eqts.~(3.11) and (3.12) are to be implemented by the equations for
 baryon and CDM density fluctuations ($\delta_b,~\delta_c$), metric
 perturbations ($\dot h_t$) and $t$ dependence of $T$. Such equations
 read:
 $$
 (\dot T/ T )^2 = \Gamma_3
 [\Omega_b + (1+w_nu) T/ T_e + \Omega_h e_h + \Omega_c] ~~,~
 $$
 $$
\dot \delta_b = -K w + 3\dot h_t/2 ~,~~~~
 \dot w = w \dot T/T + (\delta_1 - 4w)(\rho_r/\rho_b) \nu/3~,~~
 $$
 $$
 \ddot h_t = 2 \dot h_t {\dot T/ T} + \Gamma_3
 [\delta_b \Omega_b + 2 (\delta_o + \delta_{o,\nu}) T /T_e +
 e_o \Omega_h + \delta_c  \Omega_c] ~~,~$$$$
 \dot h_3/2 = - \dot h_t + (\Gamma_2/kc)
[ w \Omega_b + (\delta_1 +\delta_{1,\nu}) T/3T_e  + e_1 \Omega_h ] ~.
 \eqno (3.13)
 $$
 Here $\rho_r/\rho_b$ is the radiation/baryon density ratio, $\Omega_c T_e =
 10^4 \Omega_c T_o$ is the temperature when CDM and radiation have equal
 densities, $\Gamma_n = (8\pi/3) G \rho_c ( T /T_o )^n$, and
 $$
 e_h = \int_o^{\chi} dQ e^{-Q} \sqrt{1+Q {T^2 \over T_{de}^2} }~,~
 e_o =   \int_o^{\chi} dQ e^{-Q} {Q{T^2 \over T_{de}^2 + {1\over 2}}
 \over \sqrt{ Q+{T^2 \over T_{de}^2}} } \sigma_o ~,~
 e_1 = {T \over 2T_{de} } \int_o^{\chi} dQ e^{-Q} Q^{1/2} \sigma_1.
 \eqno (3.14)
 $$
 Here $\chi = (T_{\chi,Dy}/T)^2$ and, in most cases, it cannot be numerically
 distinguished from $\infty$.
$T_{de}$ is the typical derelativization temperature of axinos.

 We took into account also massless neutrinos. Their fluctuations are rapidly
 damped as soon as they enter the horizon. In principle, to follow such damping
 in detail, the same number of equations as for radiation are to be used. Some
 approximations, introduced when residual neutrino fluctuations are below the
 precision limit of the whole algorithm, have been set up in order to limit the
 number of harmonic components. They will not be discussed here.
In eq.(3.13), $\delta_{o,\nu}$ and
$\delta_{1,\nu}$ indicate the first two harmonics for neutrinos,
multiplied by a factor which depends on the number of massless
neutrinos in the model.

 \medskip
 \noindent
 {\bf 4. Density fluctuation evolution: numerical aspects. }

 \smallskip
 \noindent
 We performed a set of numerical integrations of the integro--differential
 system of equation (3.11), (3.12), (3.13), plus massless neutrinos. The
 integral part of the system was treated by taking 10 values of $P$ associated
 with those values of $Q$ needed to perform the integrations in eq.~(3.14) with
 the Gauss--Laguerre projection procedure. The number of $P_l$ in
 eqts.~(3.11)-(3.12) was self regulated to obtain a precision of 1:10$^5$. The
 algorithm allowed up to a maximum of 499 harmonics, which were never reached.
 The integration routine is a distant descendant of the one used by Bonometto,
 Caldara, $\&$ Lucchin (1983), although
 the treatment of collisionless particles with ntH
 spectrum required substantial changes.

 Altogether the differential equations to integrate
 numerically are up to 5505. For each model 9 lenth--scales
 were considered (1 Mpc, 5 Mpc, 10 Mpc, 20 Mpc, 50 Mpc, 100 Mpc,
 200 Mpc, 500 Mpc, 1000 Mpc), approximately spanning the
 mass--scale range from $10^{12} M_\odot$
 to $10^{21} M_\odot$. Each model requires approximately
 37 hours of CPU at a 4000/90 vaxstation. Small length--scale
 cases are obviously taking much more time than great
 length--scale ones.

 For the sake of example, in fig.~2 we report the detailed behaviour
 of density fluctuations for the various components,
 for the model 5 and the 100 Mpc case. Some irregularities are
 present in the plot of matter and radiation sound--waves,
 due to a number of prints not fully adequate to follow the
 oscillations in detail.

 In fig.~3 and 4 we plot the outcoming spectra
 at $z=0$, arising from primordial Zel'dovich spectra, for the
 models 2 and 6. In the same plots the spectra for a pure CDM
 and a standard CHDM model are also shown. The CDM spectrum has
 also been worked out with our algorithm by setting
 $\Omega_h = 10^{-3}$. The discrepancy between such spectrum
 and the one given by Holtzman (1989), for the same cosmological
 parameters, is typically $\mincir 0.5\%$, and nowhere exceeds
 $2\%$.

 In order to argue on the physical interpretation of the
 outcoming spectra, a direct comparison with CDM and CHDM
 is more suitable than the detailed spectral shape.
 Henceforth we preferred to plot the differences
 $\Delta \log[P(k)] \equiv \log[ P(k)] - \log[P(k)]_{CDM,CHDM} $,
 for the 10 cases we evaluated, rather $P(k)$ themselves.
 In fig.~4, 5 and 6 such  $\Delta \log[ P(k)]  $ are shown for the
 models 1--5, 5--7 and 8--10, respectively.

 \vfill\eject

 \noindent
 {\bf 5. Discussion. }

 \smallskip
 \noindent
 The main outputs of this work are reported in the figures
 and in Table 2, where we provide the coefficients
 for a numerical interpolation of the transfer functions
 $$
 {\cal T} (k) = \sqrt{ P(k)/k }
 = {\cal N}/\left( 1 + \sum_{j=1}^4 c_j [k/({\rm Mpc}^{-1})]^{j/2}
\right)
 \eqno (5.1)
 $$
 through the coefficients $c_j$ (the normalization coefficient
 $\cal N$ will not bew given here).

 \smallskip
 \begin{table}[tp]
 \centering
 \caption[]{Transfer function coefficients for CntHDM models}
 \tabcolsep 7pt
 \begin{tabular}{lccccccc} \\ \\
& model & $c_1 $   & $c_2 $   & $c_3 $   & $c_4 $  & \\
& 1 & -0.12991D+01 &  0.11015D+02 &  0.34827D+02 &  0.40777D+03 & \\
& 2 & -0.45687D+01 &  0.60267D+02 & -0.13429D+03 &  0.37425D+03 & \\
& 3 & -0.24013D+01 &  0.38752D+02 & -0.81353D+02 &  0.44185D+03 & \\
& 4 & -0.17309D+01 &  0.17613D+02 &  0.69582D+02 &  0.13268D+03 & \\
& 5 & -0.11754D+01 &  0.49944D+01 &  0.13379D+03 &  0.16949D+03 & \\
& 6 & -0.20006D+01 &  0.16444D+02 &  0.10363D+03 &  0.11294D+03 & \\
& 7 & -0.13807D+01 &  0.10134D+02 &  0.10300D+03 &  0.18045D+03 & \\
& 8 & -0.12392D+01 &  0.10929D+02 &  0.86743D+02 &  0.16254D+03 & \\
& 9 & -0.22635D+01 &  0.20171D+02 &  0.10958D+03 &  0.11857D+03 & \\
&10 & -0.19229D+01 &  0.11527D+02 &  0.14706D+03 &  0.15760D+03 & \\

 \end{tabular}
 \end{table}

In fig.~5, fig.~6 and fig.~7 we  compare  the spectra of
the models studied here and those of CDM and standard CHDM. Most
features reported on these figures can be qualitatively understood
taking into account the corresponding values of $\Omega_h$ and
$z_{de}$.

{\sl Model 1 (in fig. 5, solid curve).} The behaviour of this model is
not far from standard CHDM. Its transfer function nowhere exceeds
CHDM by more than a factor 1.4. No attempt was made to fine tune
$\Omega_h$ and $z_{de}$, to obtain a still nearer spectrum, but
a better fit is  possible.

{\sl Models 2 and 3 (in fig. 5: dashed curve and dotted curve).}
They are characterized by a greater $z_{de}$ and the same
$\Omega_h$, in respect to CHDM. Their transfer functions
are then greater than CHDM over all those scales which
enter the horizon when the hot component is already derelativized.
At large $k$, the curves fall down towards CHDM, as is to be expected.

{\sl Model 4 (in fig. 5: short--dashed dotted curve).}
This model differs from model 1 for a smaller $\Omega_h$ ($z_{de}$
is the same). It is therefore intermediate between CDM and CHDM.
The spectrum bends at the same scale as for CHDM and the transfer
function ratio is then decreasing fairly slowly, in respect to CDM.

{\sl Models 5--6--7 (in fig.6).}
Their behaviours show that a decrease of
$z_{de}$ can widely compensate an increase of
$\Omega_h$. The spectrum falls more rapidly than model 4 and begins to
 flatten at a smaller $k$.
These models were considered to explore the
parameter space with intermediate values for $z_{de}$.

{\sl Models 8--9--10 (in fig.7).}
While model 8 still reproduces a situation similar to CHDM, just with smaller
density for the hot component, the other two models
(with low $z_{de}$)
show a consistent decrease
of the spectrum if compared to CDM for small $k$\'s and a stabilization of
their
behaviours at $k/Mpc^{-1}\simeq 1$.
This is likely to cause a significant improvement in respect to CHDM for the
scales over which objects can form at high $z$.

The linear evolution of primeval fluctuations can be used to
evaluate a number of observational quantities  that will not
be fully enumerated here. Each one of them can be obtained
by performing suitable integrations over particular regions of the spectrum.
After COBE evaluation of $\delta T / T$, it can be convenient to compare
different models by requiring them to give the same result over the
COBE angular scale. This is approximately obtained here by requiring that
all spectra coincide at 500 Mpc. A more detailed integration of the
numerical outputs for radiation can be performed to improve $\delta T/T$
results; the radiative component is evaluated by our algorithm
down to $z \sim 500$, and we shall use all relevant harmonics to
give the expected $\delta T/T$, over a wide range of scales,
in a forthcoming work.

Among the advantages of CHDM in respect to CDM is the
behaviour of the spectrum over the scales yielding bulk velocities.
However, the faster decrease of CHDM also leads to predicting a
rather unadequate amount of high--redshift objects.
This can be a problem for high redshift galaxies and QSO's
(see, e.g., Pogosyan e Starobinsky 1994).
More recently, it has been pointed out that the problem
can be even more severe for objects associated to damped Ly$_\alpha$
clouds (Mo and Miralda-Escude`, 1994,
Subramanian and Padmanabhan, 1994,
Kauffmann and Charlot, 1994,
Ma and Bertschinger, 1994, see however also
Klypin et al., 1994).

In this connection we wish to draw the attention on the characteristics
of the models 5--6 and 9--10. Such choice of parameters give a decrease of
$P(k)$ (in respect to CDM) down to intermediate scales, while
the difference between CDM and such models tends then to
stabilize, instead of further decreasing as in standard CHDM.
More detailed computations are needed to verify up to which
degree this improving of outcoming spectra can ease the fit with
observational data.

There can be scarse doubts that, thanks to the presence of
an extra parameter, CntHDM models are more flexible
than CHDM ones. However such extra parameter is connected with
precise data of particle physics, which can be tested
through experiments which are either already feasible or
can be expected to be performed in a fairly nearby future.
This does not only concern the axino model discussed in sec.2, but also other
models with decaying particles.
Among them,  models in which there can
be decays of one neutrino flavour into lighter neutrinos can be of
particular relevance.
Although these decays have a more complex kinematics, the cosmological
behaviour due to the outcoming spectra can be very similar to the ones
discussed here.

\vglue 0.3truecm
ACKNOWLEDGMENTS. Elena Pierpaoli wishes to thank the QMW college of London,
where the last part of this paper was prepared. Silvio Bonometto thanks the
Institute of Astronomy of Cambridge (England), for its hospitality during the
preparation of the text of this paper. Gary Steigman and Stefano Borgani are
thanked for useful comments.

\vfill\eject

\centerline{\bf References }

\bigskip
\smallskip

Achilli S.,
Occhionero F. $\&$
Scaramella R.,
1985 ApJ {\bf 299}, 577

Bonometto S.,
Borgani S.,
Ghigna S.,
Klypin A. $\&$
Primack J.,
1993 MNRAS submitted

Bonometto S.,
Gabbiani F.$\&$
Masiero A.,
1994 Phis.Rev. D {\bf 49}, 3918

Bonometto S.,
Caldara A. $\&$
Lucchin F.,
1983 AA {\bf 126}, 377

Bonometto S.$\&$
Valdarnini R.,
1985 ApJ {\bf 299}, L71

Davis M.,
 Summers F.J.$\&$
 Schlegel D.,
1992 Nature {\bf 359}, 393

Ellis J. et al.,
1984 Nucl.Phys. {\bf B 238}, 453

Fang L.Z.,
 Li S.X.$\&$
 Xiang S.P.,
 1984  ApJ {\bf 140}, 77

Holtzman J.,
1989 ApJ Suppl. {\bf 71}, 1

Holtzman J. $\&$ Primack J.
  1993 ApJ {\bf 405}, 428

Kauffmann G. $\&$
Charlot S.,
1994, preprint (astro-ph/9402015)

Klypin A.,
Borgani S.,
Holtzman J. $\&$
Primack J.R.,
1994, ApJ, submitted

Klypin A.,
Holtzman J.,
Primack J. $\&$
Reg\"os E.,
1993 ApJ {\bf 415}, 1

Lee B.W. $\&$
Weinberg S.,
1977 Phys.Rev.Lett. {\bf 39}, 165

Liddle A.R. $\&$
Lyth D.H.,
1992 Phys.Lett. B, {\bf 139}, 346

Ma C.P. $\&$
Bertschinger E.,
1994, ApJL, in press (astro-ph/9407085 o 083)

Mo H.J. $\&$
Miralda-Escude` J.,
1994, preprint astro-ph/9403014

Nolthenius R.,
Klypin A. $\&$
Primack J.,
1994 ApJ {\bf 22}, L45

Pogosyan D.Y $\&$
Starobinsky A.A.
1993 preprint

Schaefer R.K. $\&$
Shafi Q.,
1992 Nature {\bf 359}, 199

Schaefer R.K.$\&$
Shafi Q.$\&$
Stecker F.,
1989 ApJ {\bf 347}, 575

Subramanian K. $\&$
Padmanabhan T.,
1994, preprint (astro-ph/9402006)

Taylor A.N. $\&$
Rowan--Robinson M.,
1992 Nature {\bf359}, 396

Valdarnini R. $\&$
Bonometto S.,
1985 AA {\bf 146}, 235

van Dalen T. $\&$
Schaefer R.K.,
1992 ApJ {\bf 398}, 33

\vfill\eject

\centerline{ FIGURE CAPTIONS }

\parindent = 0.truecm
\bigskip

Fig. 1 -- Particle and cosmological parameter for the $\tilde a$
model. The four lines refer to different choices of
 $m_\chi$ and $m_{top}$.
The solid lines are for $m_\chi = 30\, $GeV and
$m_{top} = 120\, $GeV. The dotted lines are for $m_\chi = 30\, $GeV and
$m_{top} = 180\, $GeV. The short dashed lines are for $m_\chi = 60\, $GeV and
$m_{top} = 120\, $GeV. The long dashed lines are for $m_\chi = 60\, $GeV and
$m_{top} = 180\, $GeV. The area between solid and long
dashed lines is therefore however allowed for a  combination
of  $m_\chi$ and $m_{top}$, in their allowed ranges.

Fig. 2 -- Examples of time evolution of the different components
(CDM, $\tilde a$'s, massless $\nu$'s, baryons, radiation);
$t_{rif} = 1.44 \times 10^{11}$s. The ordinate scale is arbitrary.

Fig. 3 -- Spectrum at z=0 for
the model 2 (see text).
Dotted lines refer to a CDM and
a CHDM model (the latter has $\Omega_c/\Omega_h
/\Omega_b = 0.6/0.3/0.1$).

Fig. 4 -- The same as fig.~3 for the model 6 (see text).

Fig. 5 -- The difference $\Delta \log[P(k)]$ between
the models 1 to 4 and CDM (lower curves) or CHDM (upper curves)
is given in function of $k$.
The solid curves refer to model 1.
The dotted curves   refer to model 2.
The dashed curves refer to model 3.
The short--dashed dotted curves  refer to model 4.

Fig. 6 -- The same as fig. 5. Here the models 5 to 7
are considered.
The solid curves refer to model 5.
The dotted curves   refer to model 6.
The dashed curves refer to model 7.

Fig. 7 -- The same as fig. 5. Here the models 8 to 10
are considered.
The solid curves refer to model 8.
The dotted curves   refer to model 9.
The dashed curves refer to model 10.

\end{document}
\centerline{\bf References }

\bigskip
\smallskip

Achilli S.,
Occhionero F. $\&$
Scaramella R.,
1985 ApJ {\bf 299}, 577

Bonometto S.,
Borgani S.,
Ghigna S.,
Klypin A. $\&$
Primack J.,
1993 MNRAS submitted

Bonometto S.,
Gabbiani F.$\&$
Masiero A.,
1994 Phis.Rev. D {\bf 49}, 3918

Bonometto S.,
Caldara A. $\&$
Lucchin F.,
1983 AA {\bf 126}, 377

Bonometto S.$\&$
Valdarnini R.,
1985 ApJ {\bf 299}, L71

Davis M.,
 Summers F.J.$\&$
 Schlegel D.,
1992 Nature {\bf 359}, 393

Fang L.Z.,
 Li S.X.$\&$
 Xiang S.P.,
 1984  ApJ {\bf 140}, 77

Holtzman J.,
1989 ApJ Suppl. {\bf 71}, 1

Holtzman J. $\&$ Primack J.
  1993 ApJ {\bf 405}, 428

Ellis J. et al.,
1984 Nucl.Phys. {\bf B 238}, 453

 e' questo Klip. et al 93?
Klypin A.,
Holtzman J.,
Primack J. $\&$
Reg\:os E.,
1993 ApJ {\bf 415}, 1

Nolthenius R.,
Klypin A. $\&$
Primack J.,
1994 ApJ {\bf 22}, L45

Lee B.W. $\&$
Weinberg S.,
1977 Phys.Rev.Lett. {\bf 39}, 165

Liddle A.R. $\&$
Lyth D.H.,
1992 Phys.Lett. B, {\bf 139}, 346

Pogosyan D.Y $\&$
Starobinsky A.A.
1993 preprint

Schaefer R.K. $\&$
Shafi Q.,
1992 Nature {\bf 359}, 199

Schaefer R.K.$\&$
Shafi Q.$\&$
Stecker F.,
1989 ApJ {\bf 347}, 575

Taylor A.N. $\&$
Rowan--Robinson M.,
1992 Nature {\bf359}, 396

Valdarnini R. $\&$
Bonometto S.,
1985 AA {\bf 146}, 235

van Dalen T. $\&$
Schaefer R.K.,
1992 ApJ {\bf 398}, 33

\vfill\eject

\centerline{ FIGURE CAPTIONS }

\parindent = 0.truecm
\bigskip

Fig. 1 -- Example of a result of the Monte Carlo analysis used to estimate the

\bye